\documentclass[prc,aps,amsmath,amssymb,nofootinbib,superscriptaddress,12pt]{revtex4-2}
\usepackage{epsfig}
\usepackage[bookmarksnumbered,bookmarksopen,colorlinks,citecolor=blue,linkcolor=blue]{hyperref}
\usepackage{amsmath}
\usepackage{amssymb}
\usepackage{float}

\begin{document}

    \title{ECC3: a global model for evaporation residue cross sections in the synthesis of superheavy nuclei}

\author{N. Wang}
\email{wangning@gxnu.edu.cn}
\affiliation{Department of Physics, Guangxi Normal University, Guilin 541004, China}
\affiliation{Guangxi Key Laboratory of Nuclear Physics and Technology, Guilin 541004, China}

\author{Y. J. Duan}
\affiliation{Department of Physics, Guangxi Normal University, Guilin 541004, China}

\author{H. Yao}
\affiliation{Department of Physics, Guangxi Normal University, Guilin 541004, China}

     \begin{abstract}
We present ECC3, an efficient model for evaporation residue (ER) cross sections in heavy-ion fusion reactions from light systems to superheavy-element synthesis. The ER cross section is factorized into capture, described by the universal empirical coupled channel (ECC2) model; compound-nucleus formation, given by the Zagrebaev-Greiner empirical Fermi function of the charge product; and survival against fission, obtained from a simplified HIVAP-type statistical-decay cascade. The capture sector is inherited unchanged from ECC2 and reproduces 348 measured capture excitation functions with a logarithmic RMSD of 0.29. A single set of model parameters for the formation and survival of compound nuclei is fitted to 60 measured ER excitation functions with the corresponding RMSD of 0.27. About 95\% of both the capture and ER systems are reproduced within a factor of four. Because the fitted parameters are constrained by light, medium-mass and superheavy systems rather than anchored to the measured superheavy region, extrapolation to unmeasured reactions is well controlled. The model also resolves the individual $xn$ channels. It reproduces the measured $3n$--$5n$ excitation functions of $^{19}$F+$^{188}$Os and the systematic charge dependence of the maximal channel cross section for $^{48}$Ca-, $^{50}$Ti- and $^{54}$Cr-induced reactions. For the three element-119 reactions $^{50}$Ti+$^{249}$Bk, $^{54}$Cr+$^{243}$Am and $^{51}$V+$^{248}$Cm, it predicts $4n$ dominance, with maximum cross sections of 13.2 fb, 2.64 fb and 1.56 fb, respectively.
\end{abstract}

     \maketitle

\newpage

\section{Introduction}

The synthesis of superheavy nuclei (SHN) stands at the forefront of low-energy nuclear physics, addressing fundamental questions about the limits of nuclear existence and the structure of matter at extreme proton and neutron numbers~\cite{Zhou_2020,Adamian_2020,Ogan15,Hof00}. Early theoretical predictions of an ``island of stability'' near the magic neutron number $N=184$~\cite{meldner1967predictions,Sobiczewski1966} sparked decades of experimental and theoretical exploration, culminating in the synthesis of elements up to $Z=118$ through heavy-ion fusion reactions~\cite{Thoennessen2013}. These reactions proceed through three sequential stages: capture of the projectile by the target, formation of a compound nucleus, and de-excitation of the excited compound nucleus through neutron evaporation and $\gamma$ emission. The evaporation residue (ER) cross section, which quantifies the probability that the compound nucleus survives fission and forms a detectable heavy residue, is the key observable characterizing the production yield and guides the design of new experiments.

Considerable progress in the synthesis of superheavy nuclei has been made on both the experimental and the theoretical fronts. Experimentally, cold-fusion reactions with Pb and Bi targets at GSI and RIKEN produced the elements up to $Z=113$~\cite{Hof00}, while the hot-fusion program with doubly magic $^{48}$Ca beams at Dubna extended the periodic table to $Z=118$ (oganesson) and populated the deformed-shell-stabilized nuclei around $Z=114$--116 (flerovium and livermorium)~\cite{Ogan15}. The synthesis of elements 119 and 120 is now the frontier, pursued at RIKEN, Dubna, GSI and Berkeley with heavier projectiles such as $^{50}$Ti, $^{51}$V and $^{54}$Cr~\cite{Gates2024,Oganessian2025PRC}, and new-generation facilities---the Dubna SHE Factory, the China Accelerator Facility for Superheavy Elements (CAFE) and the High Intensity heavy-ion Accelerator Facility (HIAF)~\cite{Zhou_2020,Huang2026CPL}, together with the upgraded RIKEN and GSI separators---are pushing the sensitivity toward the femtobarn level. Theoretically, macroscopic-microscopic and self-consistent mass models provide the masses, deformations and shell corrections that govern the fission barriers and the $\alpha$-decay and spontaneous-fission half-lives of these nuclei~\cite{WS4,Mo2014,Pei24,Wang15,Xu2008,Dong2010,Xie2026}. Dynamical models, such as the dinuclear system (DNS) model~\cite{Feng09,Feng07,Li_2018,Bao_2015,Li_2023,Wu_2025,Zhang_2025,Zhang_2018,Li_20181} and transport (time-dependent) approaches~\cite{Yao24,Sun2022}, describe the competition between capture, quasi-fission and compound-nucleus formation, supported by systematic studies of capture and barrier distributions~\cite{Wangbing,Chen23,Wen22,Wang08,Wang2025Universal,EBD2}. Statistical-decay codes such as HIVAP, GEMINI++ and KEWPIE2~\cite{HIVAP1981,Wang2008HIVAP,Gemini,EBD} describe the survival of the excited compound nucleus. Together, these tools now allow evaporation-residue cross sections to be estimated over the whole superheavy region and promising projectile-target combinations to be identified, including for elements 119 and 120~\cite{Wang12}.

The capture cross section, i.e.\ the probability that the colliding nuclei overcome the Coulomb barrier and become trapped in the nuclear potential pocket, is the essential first step for predicting the ER cross section and is most often computed within the DNS framework. The empirical coupled channel (ECC) method~\cite{Zagrebaev2001,Wangbing} is one of the standard tools for this purpose, in which the coupling of the relative motion to intrinsic degrees of freedom is accounted for by an asymmetric Gaussian barrier distribution characterized by a most probable barrier height $B_m$ and left and right widths $\Delta_1$ and $\Delta_2$. Because different implementations employ different interaction potentials, the ECC parameters---and therefore the predicted sub-barrier capture cross sections---vary significantly from one version to another. To remove this model dependence, the universal empirical coupled channel (ECC2) model was recently proposed~\cite{ECC2}, combining a modified Broglia-Winther (BW91$^\ast$) potential, an effective barrier curvature that reproduces the realistic width of the Coulomb barrier, and a deformation-dependent barrier-distribution width. The ECC2 most probable barrier heights agree with the values extracted from more than 300 measured excitation functions to within an rms deviation of 1.39~MeV, providing an accurate and unbiased capture input over the whole mass range.

The second and third stages, compound-nucleus formation and statistical de-excitation, are traditionally treated either by fully microscopic or by detailed statistical models. Time-dependent Hartree-Fock theory~\cite{Sim14,Yao24} and the standard statistical-decay codes provide detailed descriptions of specific aspects of the reaction, but they are computationally demanding and often require extensive tuning for each individual system. A systematic comparison of GEMINI++ and KEWPIE2 for super-asymmetric systems shows that both models reproduce the measured excitation functions reasonably well for compound nuclei with $Z=82$--$90$, whereas in the superheavy region $Z>100$ both deviate from experiment by more than an order of magnitude~\cite{PRC112_064612}. This motivates an alternative, efficient description of the compound-nucleus survival probability that keeps the essential physics but remains fast, globally consistent, and controlled by a small number of parameters.

Reliable predictions of the evaporation residue cross section therefore remain a formidable challenge. In the standard factorization $\sigma_{\rm ER}=\sigma_{\rm cap}\times P_{\rm CN}\times W_{\rm sur}$, the quoted uncertainty of current calculations is at least one to two orders of magnitude~\cite{wangbing21,PRC112_064612,Wang2026EBD3}. This is intrinsic to a simplified description of a chain of strongly coupled processes: each stage approximates a quantum many-body problem and relies on nuclear-structure inputs---deformation and fission barriers in particular---that are not known precisely, and the resulting errors accumulate along the chain. The formation probability is poorly constrained, since different models such as the dinuclear system (DNS) model~\cite{Feng09,Li_2018,Zhang_2025} and transport models give widely scattered values, and $P_{\rm CN}$ cannot be measured directly because quasi-fission products are difficult to separate from deep-inelastic and fusion-fission events. The survival probability is equally sensitive: a 1~MeV change of the fission barrier alters the $4n$ survival by about an order of magnitude. Moreover, because $\sigma_{\rm ER}$ for $Z\ge110$ remains below tens of picobarns, the available data cannot constrain all the parameters of $\sigma_{\rm cap}$, $P_{\rm CN}$ and $W_{\rm sur}$ simultaneously.

A further difficulty is methodological. Many existing calculations of superheavy evaporation residues, including the widely used DNS~\cite{Feng09,Li_2018,Zhang_2025}, determine or tune their parameters primarily on the already-measured superheavy systems. Because the measured superheavy cross sections are few and confined to a narrow region of the nuclear chart, such a parameterization is effectively anchored to the superheavy region, and its extrapolation toward lighter or yet-unmeasured systems---precisely the systems needed to disentangle the individual reaction stages---carries a large and poorly controlled uncertainty. It is therefore desirable to constrain the model with data covering as wide a range of reaction systems and energies as possible. In the present work the capture sector is taken unchanged from ECC2, which was itself constrained by a large compilation of measured capture excitation functions covering the entire mass range, and is verified here against 348 measured capture excitation functions (5573 energy points). The remaining parameters in $P_{\rm CN}$ and $W_{\rm sur}$---the compound-nucleus formation constants and the fission-survival level-density constants---are optimized globally on 60 measured evaporation-residue excitation functions (391 points) spanning the range from light asymmetric systems to the heaviest superheavy nuclei. Because the fitted parameters must simultaneously reproduce the light, medium-mass and superheavy regimes, they are not tied to the superheavy region, which greatly reduces the extrapolation uncertainty.

In this work, we present ECC3, which extends the ECC2 capture model~\cite{ECC2} to evaporation residue cross sections. The total cross section is obtained by summing over the partial waves $J$ as
\begin{equation}
\sigma_{\rm ER}(E_{\rm c.m.}) = \sum_J \sigma_{\rm cap}(E_{\rm c.m.},J)\,P_{\rm CN}(E^*)\,W_{\rm sur}(E^*,J).
\label{eq:er_factorization}
\end{equation}
Here $E^*=E_{\rm c.m.}+Q$ is the excitation energy of the compound nucleus, with $Q$ the reaction $Q$-value. $P_{\rm CN}$ is the compound-nucleus formation probability, and $W_{\rm sur}$ is the survival probability against fission during de-excitation. Each factor is described by a compact analytic or numerically cheap expression. The formation factor is an empirical Fermi-type function of the entrance-channel charge product $Z_1Z_2$ combined with an energy window, and the survival factor is obtained from a simplified statistical-decay cascade in which neutron evaporation competes with proton, $\alpha$, $\gamma$ and fission emission through Fermi-gas level densities. A globally fixed parameter set---the capture parameters of ECC2 together with the parameters in $P_{\rm CN}$ and $W_{\rm sur}$ fitted here---reproduces both the large body of capture data and the ER excitation functions from medium-mass to superheavy systems.

The present work is closely related to our previous empirical barrier distribution (EBD) models, EBD2~\cite{EBD2}, EBD2.2~\cite{EBD22} and EBD3~\cite{Wang2026EBD3}, from which ECC3 inherits the coefficients $c_0$, $c_1$ and $c_2$ of the barrier-distribution width, the $B_f^{\min}$ prescription for neutron-deficient compound nuclei, and the WS4 structure inputs. The capture sector, however, is substantially extended, and this is the main motivation for developing the universal ECC2 model~\cite{ECC2}. In EBD2.2 the capture step is described by a standard (symmetric) Gaussian barrier distribution, whereas most DNS calculations employ an asymmetric Gaussian distribution~\cite{Zagrebaev2001,Feng09}, as adopted in the present work. Moreover, the EBD2.2 parametrization is characterized only by the barrier height $V_B$ and the barrier radius $R_B$ and therefore cannot provide the depth $B_{\rm cap}$ of the potential pocket, which is an essential input of ECC3: it controls both the shallow-pocket suppression factor $F_{\rm DIS}$ and the deep-barrier suppression with $B_s=V_B-B_{\rm cap}$ in the formation probability, both introduced in Sec.~II. The BW91$^\ast$ potential of ECC2 supplies the barrier height, the barrier radius and the pocket depth over the whole nuclear chart.

This paper is organized as follows. Section~II presents the theoretical framework: the ECC2 capture cross section (Sec.~II.A), the compound-nucleus formation probability (Sec.~II.B), the survival probability (Sec.~II.C), the neutron-evaporation-channel cross sections (Sec.~II.D) and the determination of the parameters (Sec.~II.E). Section~III presents the systematic comparison with experimental data, from light systems to the heaviest superheavy elements. Concluding remarks are given in Sec.~IV.

\section{Theoretical Framework}

\subsection{Capture cross section}

The capture cross section is computed with the ECC2 model~\cite{ECC2}. At a given center-of-mass energy $E_{\rm c.m.}$, it is written as a partial-wave sum,
\begin{equation}
\sigma_{\rm cap}(E_{\rm c.m.}) = \frac{\pi\hbar^2}{2\mu E_{\rm c.m.}}\sum_J (2J+1)\,T(E_{\rm c.m.},J),
\label{eq:sigcap}
\end{equation}
where $\mu=A_1A_2/(A_1+A_2)$ is the reduced mass and $T(E_{\rm c.m.},J)$ is the probability that the colliding nuclei overcome the Coulomb barrier. In the single-barrier picture the barrier top is approximated by an inverted parabola and the penetration probability is given by the Hill-Wheeler formula~\cite{Wong73,Bass74},
\begin{equation}
T_{\rm HW}(E_{\rm c.m.},J;B) = \Big\{1+\exp\Big[-\frac{2\pi}{\hbar\omega}
\Big(E_{\rm c.m.}-B-\frac{\hbar^2 J(J+1)}{2\mu R_B^2}\Big)\Big]\Big\}^{-1},
\label{eq:hillwheeler}
\end{equation}
where $\hbar\omega$ is the barrier curvature and $R_B$ is the barrier radius. To incorporate the coupling between the relative motion and intrinsic processes such as dynamic deformation and nucleon transfer, the penetration probability is averaged over the barrier height,
\begin{equation}
T(E_{\rm c.m.},J) = \int f(B)\,T_{\rm HW}(E_{\rm c.m.},J;B)\,{\rm d}B,
\label{eq:T_av}
\end{equation}
with $f(B)$ an asymmetric Gaussian distribution,
\begin{equation}
f(B) = \begin{cases}
\displaystyle \frac{1}{\mathcal{N}}\exp\Big[-\Big(\frac{B-B_m}{\Delta_1}\Big)^2\Big], & B\le B_m,\\[8pt]
\displaystyle \frac{1}{\mathcal{N}}\exp\Big[-\Big(\frac{B-B_m}{\Delta_2}\Big)^2\Big], & B> B_m,
\end{cases}
\label{eq:fB}
\end{equation}
normalized as $\int f(B)\,{\rm d}B=1$ with $\mathcal{N}=\sqrt{\pi}(\Delta_1+\Delta_2)/2$. The most probable barrier height is
\begin{equation}
B_m = B_0 - 0.1\,Z^{1/3}\big(1+\eta\,Z^{1/3}\big),
\label{eq:Bm}
\end{equation}
where $\eta=|A_1-A_2|/(A_1+A_2)$ is the mass asymmetry of the entrance channel, $Z=Z_1+Z_2$ is the total charge, and the second term accounts for the slight increase of the BW91$^\ast$ barrier relative to the average barrier together with the downward shift of the most probable height caused by the asymmetric distribution.

The bare barrier $B_0$ is obtained from a modified Broglia-Winther potential, denoted BW91$^\ast$, in which a repulsive term $V_1$ is added to the original BW91 parametrization~\cite{BW91} to mimic the Pauli blocking at small internuclear distances,
\begin{equation}
V_N(R) = \frac{V_0+V_1}{1+\exp\!\big(\frac{R-R_N}{a}\big)},
\label{eq:BW91star}
\end{equation}
with
\begin{equation}
V_0 = -16\pi\frac{R_1R_2}{R_1+R_2}\,\gamma a, \qquad
V_1 = 1.34\Big(\frac{R_1R_2}{R_1+R_2}\Big)^2\Big(\frac{R_N}{R}\Big)^4 ~{\rm MeV},
\label{eq:V0V1}
\end{equation}
$R_N=R_1+R_2+0.29$~fm, $a=0.63$~fm, $\gamma=0.95(1-1.8 I_1I_2)$~MeV~fm$^{-2}$, $I_i=(N_i-Z_i)/A_i$, and $R_i=1.233A_i^{1/3}-0.98A_i^{-1/3}$~fm. The repulsive term $V_1$ leaves the barrier height almost unchanged but raises the capture-pocket minimum, bringing the pocket depth $B_{\rm cap}$ into close agreement with that obtained from the Skyrme energy-density functional. The Coulomb potential is
\begin{equation}
V_C = \begin{cases}
\displaystyle \frac{Z_1Z_2 e^2}{2(R_1+R_2)}\Big[3-\frac{R^2}{(R_1+R_2)^2}\Big], & R\le R_1+R_2,\\[8pt]
\displaystyle \frac{Z_1Z_2 e^2}{R}, & R> R_1+R_2.
\end{cases}
\label{eq:coulomb}
\end{equation}

A key ingredient of ECC2 is the effective barrier curvature. The standard curvature $\hbar\omega=\sqrt{-\frac{\hbar^2}{\mu}\partial^2V/\partial R^2}\big|_{R_0}$ neglects the long-range Coulomb tail and therefore yields a parabolic barrier that is systematically too narrow, which overestimates the sub-barrier penetration. ECC2 therefore adopts
\begin{equation}
\hbar\omega_{\rm eff} = \sqrt{-\frac{1}{2}\frac{\partial^2V}{\partial R^2}\Big|_{R_0}} ,
\label{eq:hw_eff}
\end{equation}
where the coefficient $\frac{1}{2}$~MeV~fm$^2$ is fixed for all reactions. The associated barrier width reproduces that of the realistic BW91$^\ast$ potential. This single change removes the systematic overprediction of sub-barrier capture that is common to earlier ECC versions.

Finally, the barrier-distribution widths are taken deformation-dependent, $\Delta_1 = \tfrac{3}{2}W$ and $\Delta_2 = 2\Delta_1$,
\begin{equation}
W = c_0(1+w_d) + c_1 B_m \sqrt{w_1^2+w_2^2+w_0^2},
\label{eq:widths}
\end{equation}
where $w_d=\sum_i|\beta_{2i}|A_i^{1/3}$ and $w_i=A_i^{1/3}\beta_{2i}^2/(4\pi)$ involve the quadrupole deformations $\beta_{2i}$ taken from the WS4 model~\cite{WS4} for prolate nuclei heavier than $^{16}$O, and $w_0=(B_m+Q)/c_2$ accounts for the dynamical effects associated with the excitation energy at the capture point. The coefficients $c_0=0.63$~MeV, $c_1=0.015$ and $c_2=33.0$~MeV are identical to those of the EBD2~\cite{EBD2} and EBD2.2~\cite{EBD22} models. For superheavy systems with shallow capture pockets, not all partial waves that overcome the Coulomb barrier contribute to capture; this is accounted for by the suppression factor~\cite{Wang2025Universal,Zhang_2025}
\begin{equation}
F_{\rm DIS} = \frac{1}{2}\Big[1+{\rm erf}\Big(\sqrt{B_{\rm cap}/c_3-1}\Big)\Big],
\label{eq:fdis}
\end{equation}
where $c_3=2.0$~MeV. The factor $F_{\rm DIS}$ enters the average barrier radius as $R_B=R_0 F_{\rm DIS}$, where $R_0$ is the barrier position of the BW91$^\ast$ potential. For light and intermediate systems the pocket is deep and $R_B\approx R_0$, whereas for shallow-pocket superheavy systems $R_B$ is appreciably smaller, reducing the above-barrier capture cross section.

\subsection{Compound-nucleus formation probability}

After capture, the dinuclear system may reseparate by quasi-fission before a compound nucleus is formed. The formation probability is parametrized as
\begin{equation}
P_{\rm CN}(E^*) = P_0(Z_1Z_2)\cdot W(E^*),
\label{eq:pcn}
\end{equation}
for light and hot fusion reactions. The asymptotic factor is the Zagrebaev-Greiner empirical Fermi function of the entrance-channel charge product~\cite{Zagrebaev2008},
\begin{equation}
P_0(Z_1Z_2) = \frac{1}{1+\exp[(Z_1Z_2-\zeta)/\tau]},
\label{eq:p0}
\end{equation}
with $\zeta=1519$ and $\tau=173$, which reflects the increasing quasi-fission suppression with the Coulomb repulsion of the entrance channel. The energy-dependent factor $W(E^*)$ is,
\begin{equation}
W(E^*) = \frac{1}{1+\exp[(E_{\rm thr}-E^*)/C]}\cdot
\frac{1}{1+\exp[(B_s-E_{\rm c.m.})/C]},
\label{eq:W_hot}
\end{equation}
where $C=2.593$~MeV and $E_{\rm thr}=B_{\rm fus}+Q$. The second factor is the deep-barrier suppression with $B_s=V_B-B_{\rm cap}$, where $V_B$ is the capture-barrier height (identified with the most probable barrier height $B_m$ in the calculations). When the incident energy is lower than the bottom of the capture pocket $B_s$, the barrier thickness felt by the fusing system increases sharply and the fusion probability drops rapidly. The effective fusion barrier is empirically expressed as,
\begin{equation}
B_{\rm fus} = \begin{cases}
B_s + E_{\rm def}, & Z_1+Z_2\ge 104,\\[2pt]
-Q, & \text{otherwise},
\end{cases}
\label{eq:Bfus}
\end{equation}
with $E_{\rm def}=E_{\rm def}^{(1)}+E_{\rm def}^{(2)}$ the sum of the ground-state deformation energies of the projectile and target, taken from the WS4 model~\cite{WS4}. This term accounts for the influence of the entrance-channel deformation on the fusion probability in hot-fusion reactions with deformed actinide targets. Finally, when the capture pocket has essentially vanished ($B_{\rm cap}\le 1$~MeV, i.e., below the zero-point energy of the dinuclear vibration), the fusion probability tends to zero and the reaction is excluded.

For light systems the compound nucleus is formed with almost unit probability, $P_{\rm CN}\approx1$, since capture and fusion are essentially Coulomb-barrier-controlled. For superheavy systems, in contrast, $P_{\rm CN}$ is strongly suppressed and may acquire an angular-momentum dependence: the quasi-fission hindrance grows with $J$, often parametrized by a critical angular momentum $L_{\rm cr}\approx45$--$60$ above which fusion is effectively suppressed~\cite{Zagrebaev2001}. This critical value is dynamical in origin---for the static BW91$^\ast$ potential with the centrifugal term, the capture pocket itself persists up to $L\approx100$ for the hot-fusion systems studied here. In the present work no explicit $J$ dependence is introduced in $P_{\rm CN}$---Eqs.~(\ref{eq:pcn})--(\ref{eq:W_hot}) contain $Z_1Z_2$ and $E^*$ only---because the angular-momentum selectivity of the evaporation-residue yield is already carried by the capture transmission and by the survival probability.

The capture transmission $T_L$ falls smoothly with $L$ through the centrifugal shift of the barrier, and the survival probability $W_{\rm sur}$ is strongly reduced at high $L$ by the rotational-energy shift $E_{\rm rot}(L)$ and by the $0.2$~MeV floor of the effective fission barrier. The partial-wave window $w_L=(2L+1)T_LW_A(E^*,L)$ is therefore bounded, lying at low $L$ for light systems and reaching $L\approx60$--$80$ for the heaviest ones. Over the range addressed in this work, $E_{\rm c.m.}<V_B+30$~MeV, this window remains well below both critical angular momenta discussed above, so that a $J$-independent $P_{\rm CN}$ is an appropriate effective description. The residual $J$ dependence associated with the quasi-fission dynamics above $L_{\rm cr}\approx45$--$60$ would change the peak cross section by at most $\approx35\%$, within the global accuracy of the model. The effect grows above $E_{\rm c.m.}\approx V_B+10$~MeV, where however the cross section has fallen well below the peak and the measured excitation functions of the superheavy systems contain few points. It becomes dominant only beyond the fitted window.

An independent, if indirect, check of this asymptotic formation probability is provided by fusion probabilities deduced from fission-fragment mass--energy distributions. For $^{48}$Ca\,+$\,^{238}$U the asymptotic value obtained with ECC3 agrees well with that extracted by Kozulin \textit{et al.}~\cite{Kozulin2016}, both being around $0.1$. Both quantities are averages: the experimental value is averaged over the partial waves contributing to the measured fission events and rises with the bombarding energy towards its plateau, while the model value is the asymptotic limit of the $J$-averaged probability defined by Eqs.~(\ref{eq:pcn})--(\ref{eq:W_hot}), whose energy-dependent rise is carried by the empirical window $W(E^*)$.

\subsection{Survival probability}

The survival probability against fission is obtained from a HIVAP-type statistical-decay cascade in the spirit of the Weisskopf-Ewing treatment. The excited compound nucleus de-excites by emitting neutrons, protons, $\alpha$ particles and $\gamma$ rays, competing with fission. Rather than performing a full two-dimensional $(Z,N)$ population evolution, ECC3 follows the neutron evaporation chain $A_0\to A_0-1\to\cdots$ and evaluates the branching ratios by integrating the emission widths over the available energy. Denoting by $\rho_c(\tilde{E})$ the level density of the decaying nucleus at its thermal excitation energy $\tilde{E}$, the neutron, proton and $\alpha$ widths are expressed through the common prefactor
\begin{equation}
\mathcal{P}_j(\tilde{E}) \equiv \frac{(2s_j+1)\,\mu_j}{\pi^2\hbar^2\,\rho_c(\tilde{E})},\qquad j=n,\,p,\,\alpha,
\label{eq:pref}
\end{equation}
and the fission and $\gamma$ widths are normalized as shown below. Each partial width then has the dimension of energy:
\begin{align}
\Gamma_n &= \mathcal{P}_n(\tilde{E})\int_0^{\tilde{E}-S_n-\Delta_d}\!\sigma_n(\varepsilon)\,\varepsilon\,\rho_d\!\big(\tilde{E}-S_n-\Delta_d-\varepsilon\big)\,W_{A-1}\!\big(\tilde{E}-S_n-\varepsilon,\,J-1\big)\,{\rm d}\varepsilon, \label{eq:In}\\[4pt]
\Gamma_p &= \mathcal{P}_p(\tilde{E})\int_0^{\tilde{E}-S_p-\Delta_c}\!\sigma_p(\varepsilon)\,\varepsilon\,
\rho_p\!\big(\tilde{E}-S_p-\Delta_c-\varepsilon\big)\,{\rm d}\varepsilon, \label{eq:Ip}\\[4pt]
\Gamma_\alpha &= \mathcal{P}_\alpha(\tilde{E})\int_0^{\tilde{E}-S_\alpha-\Delta_c}\!\sigma_\alpha(\varepsilon)\,\varepsilon\,
\rho_\alpha\!\big(\tilde{E}-S_\alpha-\Delta_c-\varepsilon\big)\,{\rm d}\varepsilon, \label{eq:Ia}\\[4pt]
\Gamma_\gamma &= C_\gamma\int_0^{\tilde{E}-\Delta_c}\!\varepsilon^3\,\rho_c\!\big(\tilde{E}-\Delta_c-\varepsilon\big)\,{\rm d}\varepsilon, \label{eq:Ig}\\[4pt]
\Gamma_f &= \frac{1}{2\pi\rho_c(\tilde{E})}\int_0^{\tilde{E}-B_f^{\rm eff}-\Delta_c}\!\rho_{\rm sad}\!\big(\tilde{E}-B_f^{\rm eff}-\Delta_c-\varepsilon\big)\,{\rm d}\varepsilon, \label{eq:If}
\end{align}
where $\tilde{E}=E^*-E_{\rm rot}(J)$ is the thermal excitation energy after removal of the rotational energy, and $S_n$, $S_p$ and $S_\alpha$ are the neutron, proton and $\alpha$ separation energies. The pairing shifts $\Delta_c$ and $\Delta_d$ of the parent and daughter nuclei are taken from the WS4 mass table~\cite{WS4}. The quantities $\sigma_n$, $\sigma_p$ and $\sigma_\alpha$ are the inverse-reaction cross sections, the charged-particle ones containing the barrier factor $\max(1-V_C^{(j)}/\varepsilon,0)$, and $\mu_j$ are the reduced masses. For neutrons the geometric inverse cross section $\sigma_n\simeq\pi R^2$ (with $R$ the radius of the decaying nucleus), together with $W_{A-1}=1$, makes the width of Eq.~(\ref{eq:In}) formally identical to the familiar Weisskopf expression $\Gamma_n=2m_nR^2/[\pi\hbar^2\rho_c(\tilde{E})]\int_0^{\tilde{E}-S_n-\Delta_d}\varepsilon\,\rho_d\,d\varepsilon$ employed in statistical-model analyses of superheavy nuclei~\cite{Zhang_2025}. In the present calculation these factors are not evaluated explicitly but are absorbed into the effective parameters discussed below. The coefficient $C_\gamma$ in Eq.~(\ref{eq:Ig}) is taken as $C_\gamma=1.2\times10^{-4}$~MeV$^{-2}$. The proton and $\alpha$ Coulomb barriers are $V_C^{(p)}=1.44(Z-1)/[1.2(A-1)^{1/3}]$~MeV and $V_C^{(\alpha)}=2\times1.44(Z-2)/[1.2(A-4)^{1/3}]$~MeV. The survival probability is then
\begin{equation}
W_A(E^*,J) = \frac{\Gamma_n+\Gamma_p+\Gamma_\alpha+\Gamma_\gamma}{\Gamma_n+\Gamma_p+\Gamma_\alpha+\Gamma_\gamma+\Gamma_f}.
\label{eq:Wsur}
\end{equation}
Here $W_A$ is the survival probability of the nucleus of mass number $A$, so that the survival probability of the compound nucleus entering Eq.~(\ref{eq:er_factorization}) is $W_{\rm sur}=W_{A_0}$. Equation~(\ref{eq:Wsur}) is a backward-cascade recursion rather than the standard product recursion of HIVAP, $W_A=(\Gamma_n/\Gamma_{\rm tot})\,W_{A-1}+\cdots$, in which the unweighted neutron width enters the total width $\Gamma_{\rm tot}$. Because the survival-weighted neutron width appears here in both the numerator and the denominator, Eq.~(\ref{eq:Wsur}) yields a slightly larger $W_A$ than the product form whenever the daughter survival is small. This difference is compensated by the global fit of the effective parameters $r_f$ and $r_h$, so Eq.~(\ref{eq:Wsur}) should be understood as an effective, empirically calibrated recursion.

The fission barrier is obtained with the WS4 predictions~\cite{WS4}, damped by the shell correction at finite excitation energy and reduced by the rotational energy,
\begin{equation}
B_f^{\rm eff}(E^*,J) = B_f - \delta_{\rm sh}\big(1-e^{-E^*/E_d}\big) - E_{\rm rot}(J),
\label{eq:Bfeff}
\end{equation}
truncated at $0.2$~MeV, with $E_{\rm rot}(J)=\hbar^2J(J+1)/(2\mathfrak{I}_{\rm rigid})$, where $\mathfrak{I}_{\rm rigid}$ is the rigid-body moment of inertia of a sphere (with radius $R=1.2A^{1/3}$~fm and deformation neglected), and $E_d=18.5$~MeV. The level densities entering Eqs.~(\ref{eq:In})-(\ref{eq:If}) are Fermi-gas expressions
\begin{equation}
\rho(U,a) = \frac{e^{2\sqrt{aU}}}{12\sqrt{2}\,a^{1/4}U^{5/4}},\qquad
a(U)=\frac{A}{K}\,r\Big[1+\delta_{\rm sh}\frac{1-e^{-U/E_d}}{U}\Big],
\label{eq:leveldensity}
\end{equation}
with the fissility-dependent parameter $K=6(1+x)$.
The fissility parameter $x$ is
\begin{equation}
x = \frac{E_C}{2E_S} = \frac{a_c Z^2/A^{1/3}}{2\,a_s(1-\kappa I^2)A^{2/3}},
\label{eq:fissility}
\end{equation}
with $I=(N-Z)/A$. The coefficients $a_s=17.409$, $a_c=0.7092$ and $\kappa=1.5189$ are taken from WS4~\cite{WS4}. In Eq.~(\ref{eq:leveldensity}), $U=E^*-\Delta_{\rm pair}$ is the thermal energy corrected for pairing and $r$ is the ratio of the level-density parameter to its ground-state value: $r=1$ for the ground-state density and $r=r_{\rm eff}$ for the saddle point (with the shell correction switched off there). For $Z\ge 82$ the saddle-point level-density parameter is enhanced relative to the ground state according to
\begin{equation}
r_{\rm eff} = r_f\;\min\!\Big[\max\!\Big(1,\big(x_0/x_{\rm ref}\big)^{3}\Big),\;r_h\Big],
\label{eq:rfe}
\end{equation}
with $r_f=1.133$, $r_h=1.616$ and $x_{\rm ref}=0.75$. $x_0$ denotes the classic fissility $\frac{a_c Z^2}{2a_s A}$. For $Z<82$ the enhancement is neglected and $r=1$.

The factor $W_{A-1}(\tilde{E}-S_n-\varepsilon,\,J-1)$ in Eq.~(\ref{eq:In}) is the survival probability of the daughter nucleus $(A-1)$: after emitting a neutron of kinetic energy $\varepsilon$, the evaporation chain continues only if the daughter nucleus survives against fission. In this simplified statistical-decay model, sub-barrier fission is neglected. Since the widths enter only through their ratios, the channel constants in Eqs.~(\ref{eq:In})--(\ref{eq:If})---the prefactors $\mathcal{P}_j(\tilde{E})$, which contain the reduced masses $\mu_j$ and the spin factors $2s_j+1$, and the inverse cross sections $\sigma_j(\varepsilon)$---are not evaluated explicitly but are absorbed into the effective fission parameters $r_f$ and $r_h$ of Eq.~(\ref{eq:rfe}), which should therefore be regarded as effective quantities. The shell correction $\delta_{\rm sh}$ is obtained by subtracting the shell-free liquid-drop binding energy, fitted to the WS4 mass table with a six-parameter liquid-drop form, from the WS4 binding energy, so that $\delta_{\rm sh}$ is defined continuously over the whole nuclear chart and is decoupled from the barrier table.

Three additional prescriptions improve the description of extreme compound nuclei, i.e., very neutron-deficient, very low-barrier or very high-fissility systems. First, for neutron-deficient compound nuclei---defined by $N-Z<60$ for superheavy nuclei (i.e. $Z\ge 104$), or a two-proton separation energy $S_{2p}<6.9$~MeV for medium-mass nuclei---the fission barrier is replaced by the minimum over the five isotones $N-2,\dots,N+2$,
\begin{equation}
B_f^{\min} = \min\big[B_f(N-2),\,B_f(N-1),\,B_f(N),\,B_f(N+1),\,B_f(N+2)\big],
\label{eq:bfmin}
\end{equation}
following the approach in Ref.~\cite{Wang2026EBD3}, considering the uncertainty of the fission barriers of nuclei near the proton drip line. Second, for superheavy nuclei with $B_f<4.6$~MeV the saddle-point enhancement is increased by $0.05$; third, for extremely fissile systems with $x>1$ the saddle-point level density is further multiplied by a power of the asymmetry-corrected fissility,
\begin{equation}
r_{\rm eff} \leftarrow r_{\rm eff}\,x^{3},
\label{eq:xgate}
\end{equation}
which suppresses the survival of the heaviest systems without affecting those with $Z\le 115$. This high-fissility gate is essential in the region where $x$ slightly exceeds unity: omitting it would overestimate $\sigma_{\rm ER}$ by about 40\% for the $Z=116$ systems ($^{50}$Ti+$^{242}$Pu and $^{54}$Cr+$^{238}$U) and by nearly an order of magnitude for $^{48}$Ca+$^{249}$Cf ($Z=118$).

\subsection{Neutron evaporation channel cross sections}

ECC3 additionally provides the partial cross sections $\sigma_{xn}$ for the individual neutron evaporation channels $x{\rm n}$ ($x=1,\dots,9$), whose sum is exactly the total $\sigma_{\rm ER}$. The channel shape is governed by the probability of realization $P_{xn}$ that exactly $x$ neutrons are emitted. Following the statistical model in~\cite{Zhang_2025} and the Jackson statistical-evaporation picture~\cite{Jackson1956}, $P_{xn}$ is expressed analytically in terms of the regularized lower incomplete gamma function for $x\ge2$ and by a Gaussian for $x=1$, with the compound-nucleus temperature and level-density parameter taken from the same Fermi-gas prescription; no adjustable parameter is introduced. The partial cross sections are then obtained by renormalizing $P_{xn}$ over the available channels,
\begin{equation}
\sigma_{xn}(E_{\rm c.m.}) = \sigma_{\rm ER}(E_{\rm c.m.})\,P_{xn}\Big/\sum_{y\ge1}P_{yn}.
\label{eq:sigxn}
\end{equation}

\subsection{Parameter determination}

The parameters of ECC3 are of two kinds. The capture parameters were determined in Ref.~\cite{ECC2} from a large compilation of measured capture excitation functions and are used unchanged. The parameters fitted for hot fusion reactions are the three formation-probability constants $\zeta$, $\tau$ and $C$ [Eqs.~(\ref{eq:pcn})--(\ref{eq:Bfus})] and the two fission-survival constants $r_f$ and $r_h$ [Eq.~(\ref{eq:rfe})]. They were determined by minimizing the pooled logarithmic RMSD, Eq.~(\ref{eq:rmsd}), between the calculated and measured evaporation-residue cross sections, where the deviations of all 391 experimental points are combined into a single average, each point entering with equal weight. The remaining quantities entering these expressions, such as the reference fissility $x_{\rm ref}$, are held fixed, as are the nuclear-structure inputs of the cascade---fission barriers, shell corrections, pairing shifts and separation energies---which are taken from the WS4 model~\cite{WS4}.

The optimization was performed on a combined set of 60 measured evaporation-residue excitation functions, containing 391 experimental points that fall inside the computed energy window (spanning the capture barrier, $V_B-8\le E_{\rm c.m.}\le V_B+30$~MeV) and enter the statistics. Because this set spans light, medium-mass and superheavy systems, the fitted parameters are constrained over the whole mass range rather than on the superheavy systems alone. The resulting values are quoted alongside the corresponding equations in Secs.~II.B and~II.C. Because the 348 capture functions constitute part of the dataset from which the ECC2 parameters were determined, their reproduction confirms that the capture sector inherited unchanged from ECC2 remains accurate over the full mass range. A genuinely independent validation of the capture sector is provided by the out-of-sample excitation functions discussed in Sec.~III.

\section{Results and Discussion}

\begin{figure}[htbp]
    \setlength{\abovecaptionskip}{0cm}
    \includegraphics[angle=0,width=0.95\textwidth]{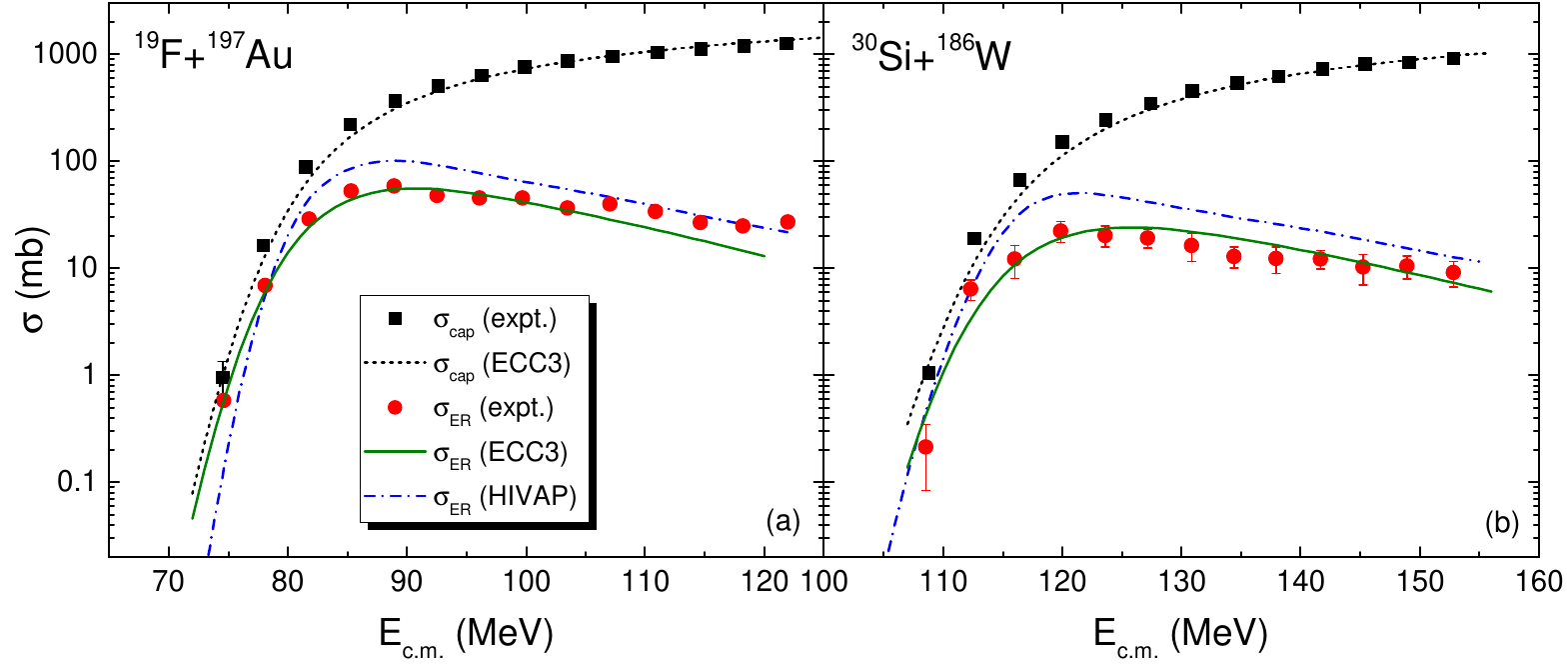}
    \caption{Predicted capture and evaporation residue cross sections for $^{19}$F\,+$\,^{197}$Au (a) and $^{30}$Si\,+$\,^{186}$W (b). The black squares and red circles denote the experimental data taken from~\cite{Hinde2002Role}. The black dotted, green solid and blue dash-dotted curves denote the ECC3 capture, ECC3 ER and HIVAP~\cite{Wang2008HIVAP} ER results, respectively.}
    \label{fig:benchmark}
\end{figure}

Figure~\ref{fig:benchmark} shows a benchmark comparison of the capture and ER description for two light asymmetric systems, $^{19}$F+$^{197}$Au and $^{30}$Si+$^{186}$W, which populate the same compound nucleus $^{216}$Ra through very different mass asymmetries. This pair tests whether the model reproduces the entrance-channel independence of a physical observable. In the upper curves the ECC3 capture cross sections (dotted curves) follow the experimental data from below the barrier to the highest measured energies, confirming the ECC2 barrier-distribution description. In the lower curves the ER cross sections are much smaller than the capture ones because of fission competition during de-excitation. The ECC3 results (solid curves) reproduce the bell shape of both excitation functions---the steep rise at the barrier and the high-energy falloff---and lie close to the HIVAP predictions (blue dash-dotted curves) for both systems. The good mutual agreement of the two systems, and of ECC3 with HIVAP, demonstrates that the simplified statistical-decay treatment is reliable for light and medium-mass systems, where fission competition is moderate and the neutron evaporation chain dominates.

\begin{figure}[htbp]
    \setlength{\abovecaptionskip}{0.2cm}
    \includegraphics[angle=0,width=1.0\textwidth]{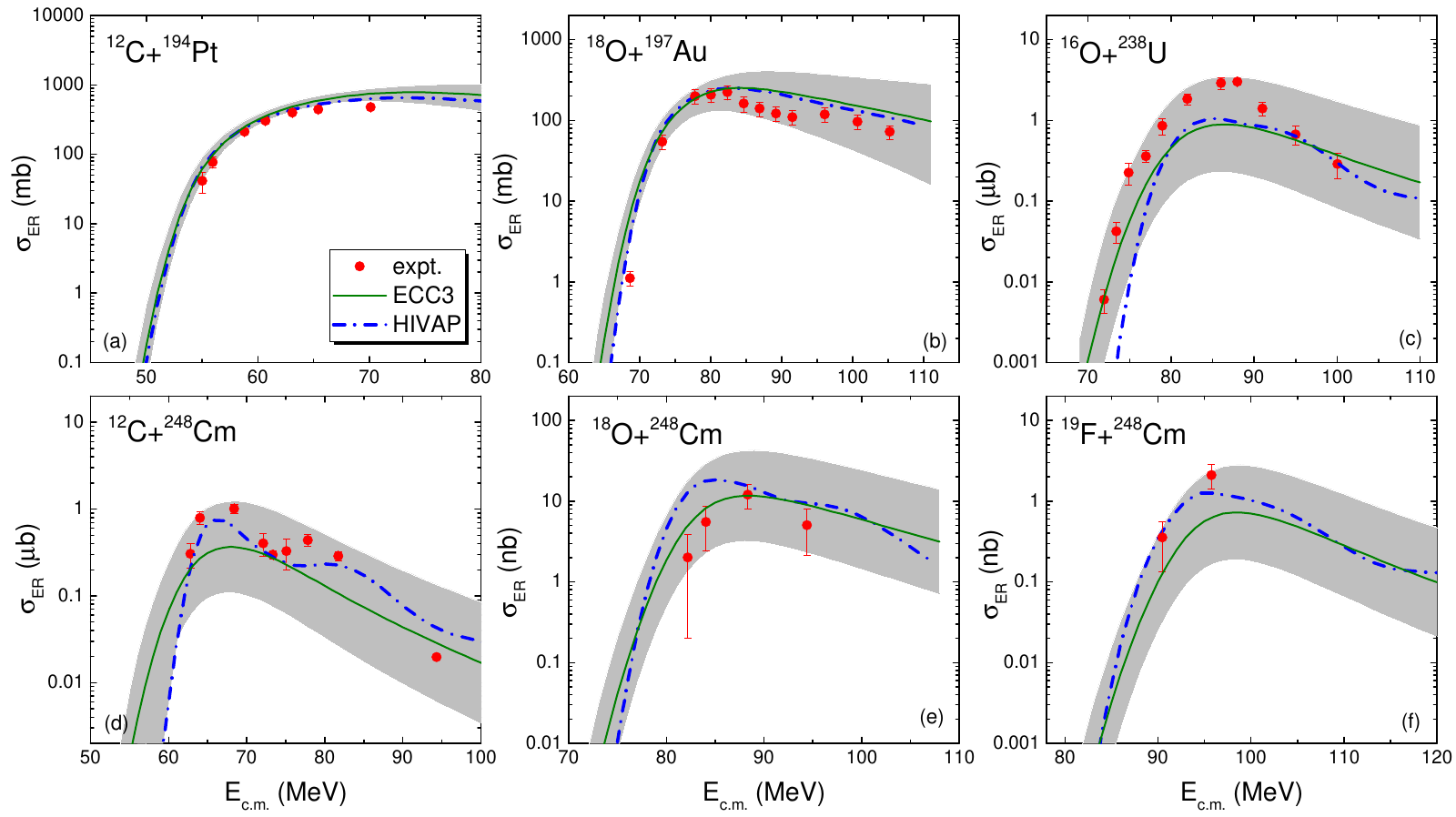}
    \caption{Predicted evaporation residue cross sections for $^{12}$C\,+$\,^{194}$Pt, $^{18}$O\,+$\,^{197}$Au, $^{16}$O\,+$\,^{238}$U, $^{12}$C\,+$\,^{248}$Cm, $^{18}$O\,+$\,^{248}$Cm and $^{19}$F\,+$\,^{248}$Cm. The red circles denote the experimental data taken from Refs.~\cite{Shrivastava1999,Gorradi2005,Nishio2004,Sikkeland1968,Murakami2013,Haba2014}. The green solid curves and the shaded bands denote the ECC3 predictions and their uncertainty, and the blue dash-dotted curves denote the HIVAP results~\cite{Wang2008HIVAP}.}
    \label{fig:systematic1}
\end{figure}

Figure~\ref{fig:systematic1} extends the comparison to six systems ordered by increasing compound-nucleus charge, from $^{12}$C+$^{194}$Pt ($^{206}$Po, $Z=84$) to $^{19}$F+$^{248}$Cm ($^{267}$Db, $Z=105$). For the lighter systems the ER cross sections reach hundreds of millibarns near the peak, so the compound nucleus predominantly survives fission and ECC3 closely follows the data. As the compound-nucleus charge increases the magnitude of $\sigma_{\rm ER}$ drops steeply---by six orders of magnitude between $^{206}$Po and $^{267}$Db---reflecting the rapid growth of fission competition and the decreasing compound-nucleus formation probability. The ECC3 curves reproduce both the magnitude and the energy dependence across the whole sequence, including the shift of the optimal energy to higher values relative to the capture barrier as fission intensifies. The comparison with HIVAP shows that the two models yield consistent results for these systems. For the heavier cases ECC3 exhibits slightly smaller deviations, owing in part to the explicit $P_{\rm CN}$ dependence on the entrance channel. The shaded bands, obtained by propagating the ECC2 capture uncertainty and the uncertainty of the saddle-point enhancement ($\pm0.05$, 1$\sigma$) in quadrature, give a realistic assessment of the predictive accuracy. For the lightest system, $^{12}$C+$^{194}$Pt, where charged-particle evaporation is largest, independent statistical-model calculations with HIVAP~\cite{Wang2008HIVAP} and GEMINI++~\cite{Gemini} give charged-particle ($pxn/\alpha xn$) contributions of only a few percent of the total ER yield even at the highest measured energy ($\approx3.5\%$ and $\approx1.9\%$ at $E^*\approx53$~MeV, respectively), so that the neutron-chain treatment of Eq.~(\ref{eq:Wsur}) remains adequate over the measured range.

\begin{figure}[htbp]
    \setlength{\abovecaptionskip}{0.2cm}
    \includegraphics[angle=0,width=1.0\textwidth]{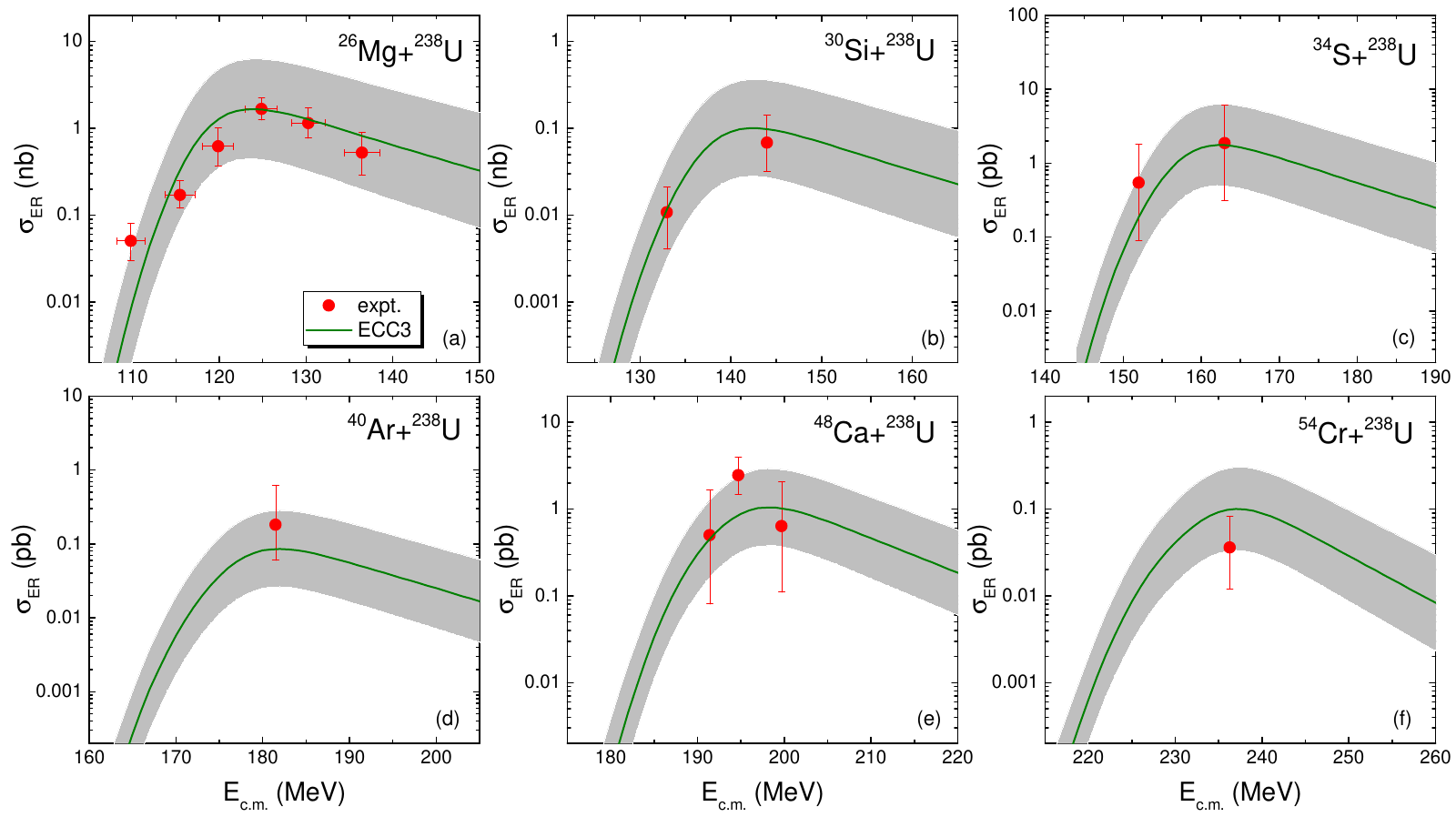}
    \caption{The same as Fig.~\ref{fig:systematic1}, but for the $^{238}$U-induced reactions $^{26}$Mg+$^{238}$U, $^{30}$Si+$^{238}$U, $^{34}$S+$^{238}$U, $^{40}$Ar+$^{238}$U, $^{48}$Ca+$^{238}$U and $^{54}$Cr+$^{238}$U (HIVAP curves are omitted). The experimental data are taken from Refs.~\cite{Gates2008,Nishio2014,Oganessian2004PRC70,Oganessian2024PRC,Oganessian2025PRC}.}
    \label{fig:systematic2}
\end{figure}

Figure~\ref{fig:systematic2} shows six reactions induced by $^{238}$U, spanning projectiles from the light $^{26}$Mg to the heavy $^{54}$Cr and compound nuclei from $^{264}$Rf ($Z=104$) to $^{292}$Lv ($Z=116$). The measured maximal cross sections decrease from the nanobarn level for $^{26}$Mg+$^{238}$U to the femtobarn level for $^{54}$Cr+$^{238}$U. ECC3 reproduces this decrease and the position and width of each excitation function with a single parameter set. The systematic trend directly reflects the combined action of the entrance-channel charge product in $P_{\rm CN}$---whose Fermi centroid $\zeta=1519$ is approached and crossed as the projectile becomes heavier---and the increase of the fission competition with $Z$. HIVAP is not compared in this figure: lacking an explicit compound-nucleus formation factor, it overestimates $\sigma_{\rm ER}$ for these heavy systems by one to two orders of magnitude because $P_{\rm CN}\ll1$, whereas ECC3 accounts for this suppression through $P_{\rm CN}$. For the heaviest system, $^{54}$Cr+$^{238}$U, the cross section reaches only tens of femtobarns, at the present sensitivity limit of superheavy-element experiments, yet the ECC3 prediction remains consistent with the single available data point. The agreement across this figure demonstrates that the model captures both the quasi-fission suppression and the statistical-decay competition in the superheavy regime.

\begin{figure}[htbp]
    \setlength{\abovecaptionskip}{0.5cm}
    \includegraphics[angle=0,width=1\textwidth]{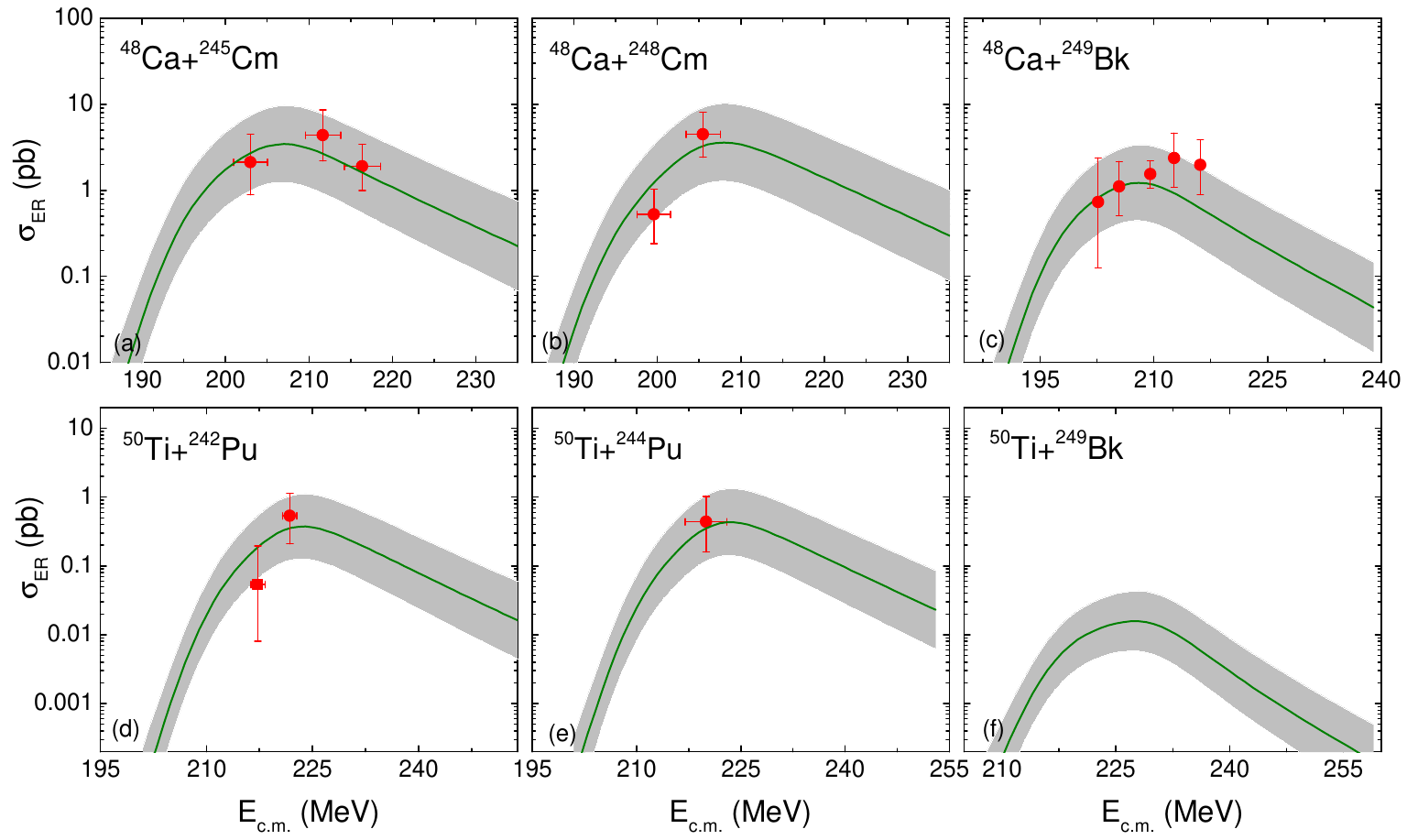}
    \caption{Predicted evaporation residue cross sections for the hot-fusion reactions $^{48}$Ca+$^{245,248}$Cm, $^{48}$Ca+$^{249}$Bk, $^{50}$Ti+$^{242,244}$Pu and $^{50}$Ti+$^{249}$Bk. The red circles denote the experimental data taken from Refs.~\cite{Oganessian2004PRC69,Oganessian2015PRC92,Oganessian2015RPP,Oganessian2025PRC,Oganessian2026PRC,Gates2024}; the shaded bands and green solid curves denote the ECC3 predictions and their uncertainty.}
    \label{fig:systematic3}
\end{figure}

Figure~\ref{fig:systematic3} focuses on hot-fusion reactions leading to the heaviest known elements and to element 119, with compound nuclei from $^{293}$Lv ($Z=116$) to the element-119 system $^{50}$Ti+$^{249}$Bk. The measured peak cross sections are at the picobarn level. ECC3 reproduces the peak positions, widths and magnitudes of the $^{48}$Ca-induced reactions within the experimental uncertainties. For the $^{50}$Ti-induced reactions the ECC3 optimal energies shift to higher values, following the increased Coulomb repulsion of the heavier projectile, in agreement with the data. The result for $^{50}$Ti+$^{249}$Bk, where no data are yet available, illustrates the predictive use of the model: ECC3 places the maximum of the excitation function near $E_{\rm c.m.}\approx 229$~MeV with a peak cross section of about 13 fb, driven by the $4n$ evaporation channel. The overall agreement across this figure shows that a single, globally fixed parameter set can describe both the $^{48}$Ca and the heavier $^{50}$Ti projectile families.

\begin{figure}[htbp]
    \setlength{\abovecaptionskip}{0.2cm}
    \includegraphics[angle=0,width=0.85\textwidth]{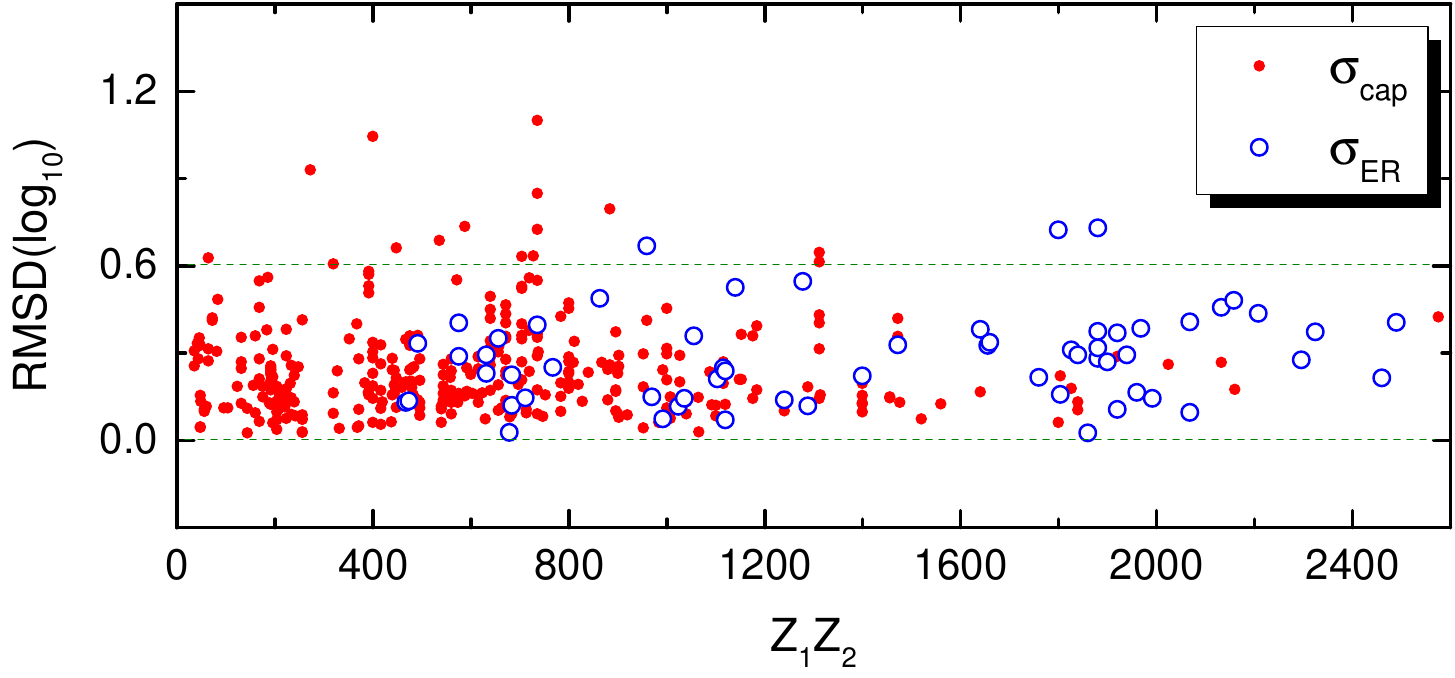}
    \caption{Logarithmic RMSD of the ECC3 capture cross sections (filled red circles; 348 measured excitation functions) and of the ECC3 evaporation residue cross sections (open blue circles; 60 fusion systems from light to superheavy) as a function of the entrance-channel charge product $Z_1Z_2$. The green dashed lines mark RMSD$=0$ and RMSD$=0.6$. About 95\% of all systems fall below the 0.6 line, corresponding to an average deviation factor of four.}
    \label{fig:rmsd}
\end{figure}

To quantify the global performance, Fig.~\ref{fig:rmsd} shows the logarithmic RMSD,
\begin{equation}
{\rm RMSD} = \sqrt{\frac{1}{n}\sum_{i=1}^{n}
\Big[\log_{10}\sigma^{\rm exp}_i-\log_{10}\sigma^{\rm theo}_i\Big]^2},
\label{eq:rmsd}
\end{equation}
for the capture cross sections of 348 measured excitation functions (filled circles) and for the ER cross sections of 60 fusion systems from light to superheavy (open circles), plotted against the entrance-channel charge product $Z_1Z_2$. The capture distribution is centred at RMSD$\approx0.24$ with a pooled value of 0.29 over 5573 data points, and the ER distribution at RMSD$\approx0.28$ with a pooled value of 0.27 over 391 data points. Each symbol in Fig.~\ref{fig:rmsd} represents the per-system RMSD obtained from Eq.~(\ref{eq:rmsd}) with $n$ the number of experimental points of that system, so that every system enters the distribution with equal weight. The quoted pooled values, in contrast, are obtained by applying Eq.~(\ref{eq:rmsd}) to the deviations of all points of all systems combined, so that every data point carries equal weight. No systematic drift with $Z_1Z_2$ is visible: the model performs equally well for light fusion systems and for the heaviest superheavy reactions. Roughly 95\% of both the capture and the ER systems fall below the 0.6 line, i.e., they are reproduced to within a factor of four. This global accuracy is the central result of the present work: it spans 348 capture and 60 evaporation-residue systems and nearly 6000 experimental energy points, with the capture sector inherited unchanged from ECC2 and only the parameters in $P_{\rm CN}$ and $W_{\rm sur}$ fitted to the ER data. It should be emphasized that the fitted parameters are constrained simultaneously by light, medium-mass and superheavy systems, so that the model is not anchored to the measured superheavy region and its extrapolation to unmeasured systems is comparatively well controlled. A systematic test of this kind, covering the entire mass range with one fixed parameter set, is not usually performed for the DNS models. Their parameters are adjusted mainly on the measured superheavy systems. The present level of accuracy therefore appears to be among the best currently achieved by models of this class, while remaining computationally inexpensive.

Since the fission barriers are the most influential input of the survival cascade, we also examined the sensitivity of the results to the barrier model. Replacing the WS4 barriers for the $Z=114$ and $Z=116$ isotopic chains by the macroscopic-microscopic values of Kowal, Jachimowicz and Sobiczewski~\cite{Kowal2010} raises the peak evaporation-residue cross sections by $33\%$--$56\%$ ($+0.12$ to $+0.19$~dex) and moves the maxima to about $2$~MeV lower excitation energy. The Kowal et al.\ barriers are higher than the WS4 ones at the compound nuclei by $0.75$--$1.01$~MeV, and the cross-section increase grows with this barrier difference: $+33\%$ for $^{48}$Ca+$^{242}$Pu ($^{290}$Fl, $\Delta B_f=+0.75$~MeV), $+52\%$ for $^{48}$Ca+$^{244}$Pu ($^{292}$Fl, $\Delta B_f=+1.01$~MeV) and $+56\%$ for $^{50}$Ti+$^{242}$Pu ($^{292}$Lv, $\Delta B_f=+0.89$~MeV, whose whole $Z=116$ chain is shifted upwards). Because even the largest shift ($\approx0.19$~dex) remains below the global accuracy of the model ($0.27$~dex), the present parametrization is not critically tied to the choice of barrier model.

The extrapolation capability of the model can also be tested directly against data that were not included in the fit. For the capture sector, twelve excitation functions (195 points) for systems ranging from $^{11}$B+$^{12}$C and $^{19}$F+$^{12}$C to $^{11}$B+$^{235}$U, $^{11}$B+$^{238}$U and $^{48}$Ca+$^{244}$Pu---none of which belongs to the 348 functions of Fig.~\ref{fig:rmsd}---give a pooled logarithmic RMSD of 0.18, with per-system values between 0.07 and 0.43. For the ER sector, four new excitation functions---$^{30}$Si+$^{170}$Er~\cite{Hinde1982}, $^{32}$S+$^{184}$W~\cite{Back1999}, $^{40}$Ar+$^{181}$Ta~\cite{Vermeulen1984} and $^{48}$Ti+$^{122}$Sn~\cite{Gil1995}---none of which belongs to the 60 functions of Fig.~\ref{fig:rmsd}, give a pooled RMSD of 0.21, with per-system values between 0.14 and 0.42. Both out-of-sample values are smaller than the RMSD obtained on the fit sets themselves (0.29 and 0.27), showing that the extrapolation to systems outside the fitted dataset is indeed well controlled.

\begin{figure}[htbp]
    \setlength{\abovecaptionskip}{0.2cm}
    \includegraphics[angle=0,width=1.0\textwidth]{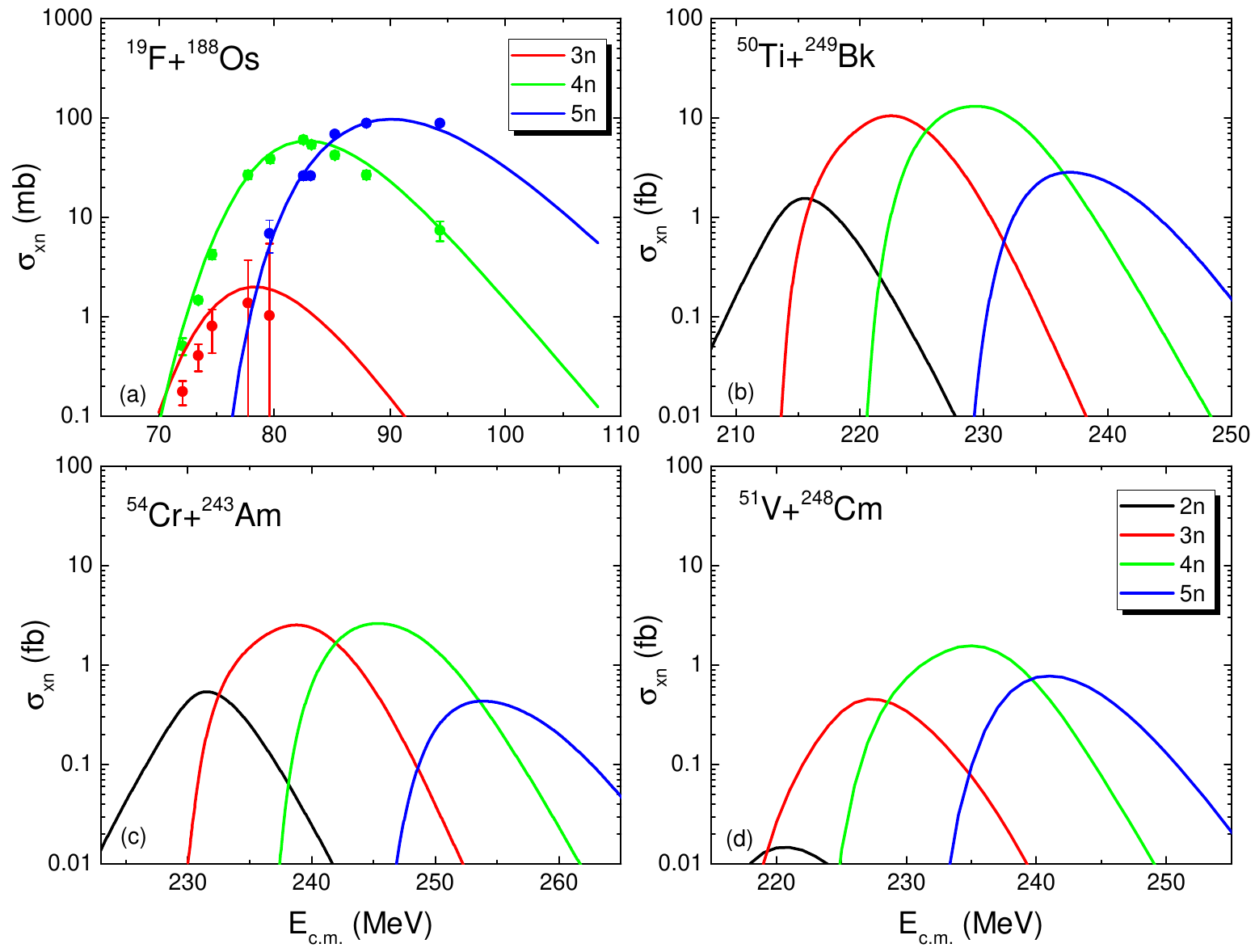}
    \caption{Neutron evaporation channel cross sections $\sigma_{xn}$ predicted by ECC3. (a) $^{19}$F+$^{188}$Os, showing the $3n$, $4n$ and $5n$ channels; the symbols denote the experimental data~\cite{Mahata2003}. (b)--(d) the $2n$--$5n$ channels for the three reactions synthesizing element 119, $^{50}$Ti+$^{249}$Bk, $^{54}$Cr+$^{243}$Am and $^{51}$V+$^{248}$Cm.}
    \label{fig:xn}
\end{figure}

Figure~\ref{fig:xn} demonstrates that ECC3 resolves the total ER cross section into individual neutron evaporation channels. Panel (a) compares the $3n$, $4n$ and $5n$ excitation functions of $^{19}$F+$^{188}$Os with the measured data~\cite{Mahata2003}; the model reproduces the positions and the relative magnitudes of the three channels, with the maximum in the $5n$ channel. Panels (b)--(d) show the predicted $2n$--$5n$ channels for the three reactions that lead to element 119, $^{50}$Ti+$^{249}$Bk, $^{54}$Cr+$^{243}$Am and $^{51}$V+$^{248}$Cm, at the femtobarn level. For all three systems the $4n$ channel dominates, and the channel maxima shift to higher energies as the projectile becomes heavier, in line with the increasing Coulomb barrier. The predicted $4n$ maxima are 13.2, 2.64 and 1.56~fb for $^{50}$Ti+$^{249}$Bk, $^{54}$Cr+$^{243}$Am and $^{51}$V+$^{248}$Cm, located at optimal energies $E_{\rm c.m.}=229$, 245 and 235~MeV, respectively. The somewhat smaller $4n$ maximum of $^{51}$V+$^{248}$Cm compared with $^{54}$Cr+$^{243}$Am is mainly a survival effect: the two reactions populate different isotopes of element 119 ($^{299}$119 and $^{297}$119), and the optimal excitation energy of $^{51}$V+$^{248}$Cm is about $2$~MeV higher ($E^*\approx40$~MeV versus $\approx38$~MeV), where fission competes more strongly and the survival probability is correspondingly reduced. Varying the fission-saddle level-density enhancement $r_{\rm eff}$ by $\pm0.05$ gives asymmetric theoretical uncertainties of $^{+22.3}_{-8.2}$, $^{+4.5}_{-1.7}$ and $^{+2.8}_{-1.0}$~fb---about a factor of 2.7 in each direction, consistent with the factor-of-four global accuracy of Fig.~\ref{fig:rmsd}. The channel shapes are smooth, well-separated peaks, a consequence of the narrow multiplicity distribution of the probability of realization. These results show that, in addition to the total ER cross section, ECC3 provides channel-resolved predictions directly comparable with the isotopic yields measured in superheavy-element experiments.

\begin{figure}[htbp]
    \setlength{\abovecaptionskip}{0.2cm}
    \includegraphics[angle=0,width=0.65\textwidth]{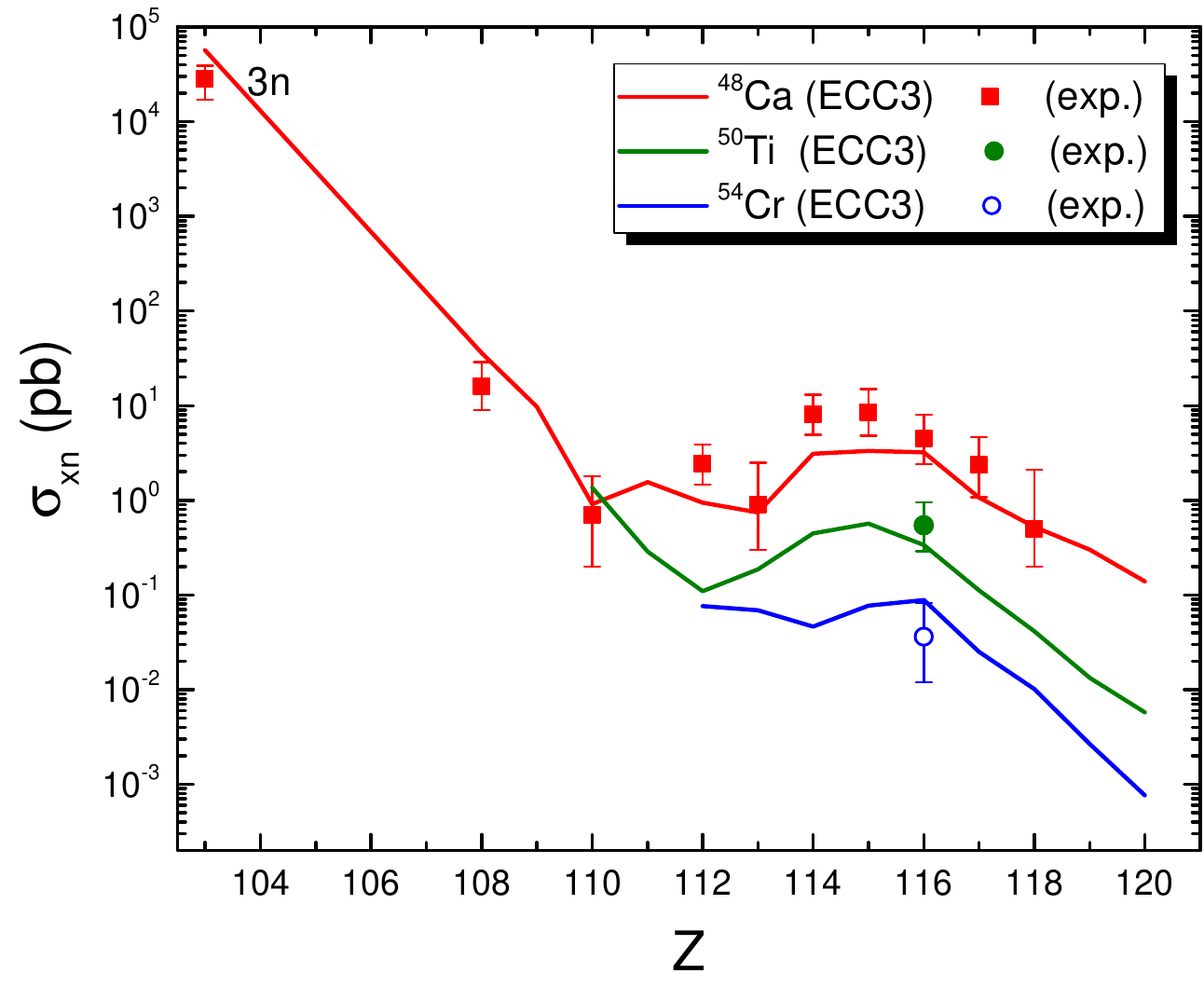}
    \caption{Systematic dependence of the maximal evaporation-channel cross section $\sigma_{xn}$ on the compound-nucleus charge $Z$ for the $^{48}$Ca- (red), $^{50}$Ti- (green) and $^{54}$Cr-induced (blue) hot-fusion reactions. For each reaction, $\sigma_{xn}$ denotes the peak cross section of the dominant neutron evaporation channel, obtained by maximizing $\sigma_{xn}(E_{\rm c.m.})$ over the channels $x$ and over the energy. The curves denote the ECC3 predictions and the symbols the experimental values. The targets used are $^{209}$Bi, $^{226}$Ra, $^{227}$Ac, $^{232}$Th, $^{231}$Pa, $^{238}$U, $^{237}$Np, $^{242}$Pu, $^{243}$Am, $^{248}$Cm, $^{249}$Bk, $^{249}$Cf, $^{252}$Es and $^{258}$Fm.}
    \label{fig:systematics_z}
\end{figure}

Finally, Fig.~\ref{fig:systematics_z} summarizes the systematic behaviour of the maximal evaporation-channel cross section $\sigma_{xn}$---the peak cross section of the dominant neutron evaporation channel, maximized over the channels $x$ and over the energy---as a function of the compound-nucleus charge $Z$ for the three hot-fusion projectile families $^{48}$Ca, $^{50}$Ti and $^{54}$Cr, using a target series from $^{209}$Bi to $^{258}$Fm. The ECC3 curves reproduce the overall trend, including the local maximum (a pronounced shoulder) of the $^{48}$Ca-induced reactions in the region of the deformed-shell-stabilized nuclei around $Z\approx114$--116 (flerovium and livermorium), where the fission barrier is restored by shell effects, followed by a steep decrease as $Z$ grows towards 118. Moreover, the systematic ordering is correctly captured: the $^{54}$Cr-induced cross sections lie noticeably below the $^{50}$Ti and $^{48}$Ca ones, reflecting both the larger entrance-channel charge product (stronger quasi-fission suppression through $P_{\rm CN}$) and the higher fissility of the heavier compound nuclei (weaker survival). The comparison confirms that ECC3 reproduces the charge dependence of the ER yield over the full superheavy region and provides a consistent framework for extrapolating to yet-unobserved isotopes.

\section{Summary}

We have presented ECC3, an efficient model for the systematic description of evaporation residue cross sections in heavy-ion fusion reactions from light asymmetric systems to superheavy-element synthesis. The total ER cross section is factorized into three stages: capture, computed with the ECC2 empirical coupled channel model (BW91$^\ast$ potential, effective barrier curvature and deformation-dependent asymmetric Gaussian barrier distribution); compound-nucleus formation, described by an empirical Fermi function of the entrance-channel charge product combined with an energy window; and statistical de-excitation, treated by a simplified cascade along the neutron evaporation chain with competing neutron, proton, $\alpha$, $\gamma$ and fission widths evaluated with Fermi-gas level densities. Fission-barrier shell damping, a saddle-point level-density enhancement controlled by the compound-nucleus fissility, the $B_f^{\min}$ treatment for neutron-deficient compound nuclei and a superheavy gate for $x>1$ complete the model. The capture sector is taken unchanged from ECC2, which was constrained by a large compilation of measured capture excitation functions, and only the parameters---the compound-nucleus formation constants and the fission-survival level-density constants---are fitted here, to 60 measured evaporation-residue excitation functions spanning light, medium-mass and superheavy systems. Because the fitted parameters are not anchored to the measured superheavy region, the extrapolation to lighter and to yet-unmeasured systems is comparatively well controlled.

ECC3 has been validated against 348 measured capture excitation functions and 60 fusion systems from light to superheavy. The pooled logarithmic RMSD is 0.29 for capture and 0.27 for the ER cross sections; about 95\% of the systems are reproduced to within a factor of four, without systematic drift with the entrance-channel charge product. This level of accuracy is comparable to or better than that of existing models of the same class, while the computation remains inexpensive. The model reproduces the ER excitation functions of benchmark light systems, the six-system sequence from $^{206}$Po to $^{267}$Db (consistent with HIVAP), the $^{238}$U-induced reactions from nanobarns to femtobarns, and the picobarn-level hot-fusion reactions of the $^{48}$Ca and $^{50}$Ti families. It further provides channel-resolved $xn$ cross sections, reproducing the measured $3n$--$5n$ channels of $^{19}$F+$^{188}$Os and predicting a $4n$ dominance for the element-119 reactions, and it reproduces the systematic dependence of the maximal channel cross section on the compound-nucleus charge for $^{48}$Ca-, $^{50}$Ti- and $^{54}$Cr-induced reactions.

These results establish ECC3 as a fast, globally consistent and few-parameter tool for estimating evaporation residue cross sections and guiding experimental searches for new isotopes. Because the capture and survival sectors use globally fixed parameter sets that are not anchored to the measured superheavy region, the model can be applied immediately to unmeasured projectile-target combinations, including the Ti-, Cr- and V-induced reactions currently under investigation. Online calculations are available at \url{http://www.imqmd.com/fusion/ECC3/}.

\section*{Acknowledgments}
This work was supported by the Guangxi ``Bagui Scholar'' Teams for Innovation and Research Project, the Guangxi Natural Science Foundation (No. 2026GXNSFBA00640195) and the National Natural Science Foundation of China (Nos. 12265006, U1867212, 12475121). We are grateful to S.-G. Zhou and B. Wang for providing their codes for the statistical-decay calculations. Helpful discussions with members of the Superheavy Nuclei Theory Collaboration (SNTC) are highly appreciated.

\section*{Data Availability Statement}
The data that support the findings of this study, including the compiled experimental datasets and the ECC3 program, are available from the corresponding author upon reasonable request and at \url{http://www.imqmd.com/fusion/}.

\end{document}